\documentclass[conference]{IEEEtran}
\usepackage[hyphens]{url}
\usepackage{hyperref}
\hypersetup{breaklinks=true,colorlinks,allcolors=blue}
\usepackage{doi}

\usepackage[backend=bibtex,style=ieee,citestyle=numeric-comp]{biblatex}
\usepackage{booktabs}
\usepackage{amsmath,amssymb,amsfonts}
\usepackage{algorithmic}
\usepackage{graphicx}
\usepackage{textcomp}
\usepackage{xcolor}
\usepackage{balance}
\usepackage{enumitem} 
\def\BibTeX{{\rm B\kern-.05em{\sc i\kern-.025em b}\kern-.08em
    T\kern-.1667em\lower.7ex\hbox{E}\kern-.125emX}}

\begin{document}

\title{The Need for Standardized Evidence Sampling in CMMC Assessments: A Survey-Based Analysis of Assessor Practices}

\author{\IEEEauthorblockN{Logan Therrien}
\IEEEauthorblockA{\textit{The Beacom College of Computer and Cyber Sciences} \\
\textit{Dakota State University}\\
Madison, United States \\
logan.therrien@trojans.dsu.edu}
\and
\IEEEauthorblockN{John Hastings}
\IEEEauthorblockA{\textit{The Beacom College of Computer and Cyber Sciences} \\
\textit{Dakota State University}\\
Madison, United States \\
john.hastings@dsu.edu}
}

\maketitle

\begin{abstract}
The Cybersecurity Maturity Model Certification (CMMC) framework provides a common standard for protecting sensitive unclassified information in defense contracting. While CMMC defines assessment objectives and control requirements, limited formal guidance exists regarding evidence sampling, the process by which assessors select, review, and validate artifacts to substantiate compliance. Analyzing data collected through an anonymous survey of CMMC-certified assessors and lead assessors, this exploratory study investigates whether inconsistencies in evidence sampling practices exist within the CMMC assessment ecosystem and evaluates the need for a risk-informed standardized sampling methodology. Across 17 usable survey responses, results indicate that evidence sampling practices are predominantly driven by assessor judgment, perceived risk, and environmental complexity rather than formalized standards, with formal statistical sampling models rarely referenced. Participants frequently reported inconsistencies across assessments and expressed broad support for the development of standardized guidance, while generally opposing rigid percentage-based requirements. The findings support the conclusion that the absence of a uniform evidence sampling framework introduces variability that may affect assessment reliability and confidence in certification outcomes. Recommendations are provided to inform future CMMC assessment methodology development and further empirical research.
\end{abstract}

\begin{IEEEkeywords}
CMMC, evidence sampling, cybersecurity assessments, audit methodology, assessor judgment, inter-assessor reliability, C3PAO
\end{IEEEkeywords}

\section{Introduction}
The Cybersecurity Maturity Model Certification (CMMC) program was developed by the U.S. to improve the protection of Controlled Unclassified Information (CUI) within the Defense Industrial Base (DIB) \cite{DoD_CMMC_32CFR170_2024_b1}. CUI is defined by the U.S. government as information requiring safeguarding or dissemination controls pursuant to 32 CFR Part 2002 \cite{NARA_CUI_32CFR2002_2016_b2}. These CUI protection obligations are commonly flowed to defense contractors through DFARS 252.204-7012 \cite{DFARS_252_204_7012_b3} and associated DFARS clauses \cite{DFARS_252_204_7019,DFARS_252_204_7020,DFARS_252_204_7021}. CMMC Level 2 relies on independent, third-party assessments conducted by accredited Certified Third-Party Assessment Organizations (C3PAOs) and CMMC certified assessors \cite{DoD_CMMC_32CFR170_2024_b1,b5}. The assessment objectives are aligned to the security requirements in NIST SP 800-171 \cite{NIST_SP800_171r2_2020_b6}. The procedures for evaluating implementation are commonly operationalized using assessment-method guidance such as NIST SP 800-171A \cite{NIST_SP800_171A_2018_b4}.
This structure places significant emphasis on assessment rigor, repeatability, and fairness.

While CMMC defines control requirements and assessment objectives, it provides comparatively limited prescriptive guidance on how evidence should be sampled to validate those requirements \cite{DoDCIO_CMMC_AssessmentGuideL2_2024, CyberAB_CAP_v2_2024, NIST_SP800_171r2_2020_b6}. For CMMC Level 2 assessments, scoping guidance distinguishes in-scope assets and environments from out-of-scope elements and provides assessment-relevant definitions that influence what evidence is reasonably available for sampling \cite{DoDCIO_CMMC_ScopingGuideL2_v2_2024}. Evidence sampling is a critical component of any assessment process, as it directly influences the confidence an assessor can place in conclusions drawn about system compliance \cite{NIST_SP800_171A_2018_b4}. Inconsistent sampling approaches may lead to divergent outcomes for organizations with similar security postures, undermining stakeholder trust in certification results.

In established auditing and assessment disciplines, evidence sampling is treated as a formal methodological concern, often guided by risk-based principles, representativeness requirements, and documentation standards \cite{IAASB_ISA530_2009,ISOIEC_27007_2020,Hennink_Kaiser_Saturation_2022_b7}. 
In cybersecurity and compliance contexts, risk-based decisions are commonly operationalized by assessing likelihood and impact to prioritize the depth of evaluation and the assurance needed from collected evidence \cite{NIST_SP800_30r1_2012}. In contrast, anecdotal observations within the CMMC ecosystem suggest that sampling decisions are frequently left to assessor discretion or internal C3PAO practices. This study seeks to empirically examine these observations by analyzing assessor-reported practices and perspectives. The main contributions of this research are an empirical snapshot of variability in CMMC evidence sampling, a scenario-based demonstration of assessor dispersion, and a synthesis of practitioner-supported directions for standardizing sampling guidance. The research is guided by the following research questions:

\begin{enumerate}[label={\textbf{RQ\arabic*:}},left=1.0em]
    \item What factors drive sampling decisions (risk, complexity, judgment)?
    \item How much dispersion appears under identical scenarios?
    \item How commonly do assessors experience inconsistency across assessments/C3PAOs?
    \end{enumerate}

\section{Motivation}
This research problem was identified through sustained practitioner observation within the CMMC assessment ecosystem. Through direct participation in assessments, professional collaboration with other assessors, and exposure to multiple C3PAO practices, recurring inconsistencies were observed in how evidence was sampled, evaluated, and deemed sufficient to support requirement implementation. These observations revealed that assessors often applied divergent sampling thresholds and expectations under similar conditions, leading to assessment friction, extended timelines, and disagreement regarding compliance sufficiency. The persistence of these observations across organizations and engagements indicated a systemic issue rather than isolated assessor behavior, motivating the formal investigation of evidence sampling as a methodological gap within the CMMC assessment framework.

\section{Methodology}

The research adopts a practical mixed-methods approach, combining quantitative and qualitative data to best address the research problem, assessor variability in evidence sampling, which may not be adequately understood through purely quantitative or qualitative means alone \cite{Creswell_PlanoClark_MixedMethods_2018}. Quantitative data provides insight into patterns of experience and responses to structured scenarios, while qualitative data captures contextual reasoning, professional judgment, and perceptions of inconsistency. 
The study is exploratory and diagnostic in nature, appropriate given the limited existing empirical research on CMMC evidence sampling practices. 

\subsection{Survey Design}
Design of a structured survey instrument targeting CCAs and LCCAs. The survey instrument used in this study was intentionally developed to address the absence of empirical research examining evidence sampling practices within the CMMC assessment ecosystem. Survey questions were derived through an iterative design process informed by:

\begin{enumerate}
    \item CMMC assessment objectives and procedures
    \item Established audit and assessment sampling principles, and
    \item Practitioner observations regarding variability in assessor expectations

\end{enumerate}

To ensure relevance and construct coverage, questions were mapped directly to core dimensions of evidence sampling, including assessor background and experience, decision drivers for sample selection, application of sampling in structured scenarios, perceptions of consistency across assessments, and professional perspectives on standardization. This design approach ensured that each question served a specific analytical purpose aligned with the study's research questions rather than functioning as a general opinion poll.

The instrument intentionally combined structured response items with open-ended questions to support a mixed-methods analytical strategy. Structured items enabled comparative analysis across respondents, while narrative prompts captured contextual reasoning and professional judgment that cannot be adequately represented through fixed response options alone. Scenario-based questions were incorporated to elicit applied decision-making under controlled conditions, reducing ambiguity and allowing for direct comparison of assessor approaches. Content validity was further strengthened by framing questions in operational assessment language familiar to CCAs and LCCAs. Collectively, the survey questions were designed to balance rigor with practical relevance, enabling systematic analysis of evidence sampling practices while preserving the depth necessary to identify methodological gaps within the current CMMC assessment framework.

\subsection{Data Collection \& Ethical Considerations}
Data collection was conducted through a web-based survey designed to protect participant confidentiality and encourage candid responses \cite{Dillman_Smyth_Christian_2014}. Anonymity was a critical design consideration given the professional sensitivity of CMMC assessments; therefore, no personally identifiable information was collected, and IP addresses were not observed or retained as part of the data collection process. Survey invitations were distributed via email exclusively to individuals identified as CCAs and LCCAs listed in the official CyberAB marketplace \cite{CyberAB_Marketplace_Catalog}, ensuring that participation was restricted to the intended professional population. This targeted and anonymous approach reduced response bias, increased the likelihood of honest disclosure regarding assessment practices, and strengthened the validity of the collected data. Ethical oversight was ensured through institutional IRB DSUIRB-20251118-03EX.

\subsection{Data Cleaning}
Collected data underwent cleaning and descriptive profiling to ensure the accuracy, usability, and interpretability of the collected survey responses prior to analysis. Raw survey data was reviewed to remove incomplete records and non-informative fields. Descriptive profiling was then conducted to characterize the dataset, including respondent counts, the distribution of response types, and the identification of open-ended versus structured items. This process enabled the validation of dataset integrity, informed the selection of appropriate analytical techniques, and ensured that subsequent qualitative and quantitative analyzes were grounded in a clean, well-understood data foundation.

\subsection{Thematic Coding}
Open-ended survey responses were thematically coded to systematically analyze qualitative data related to evidence sampling practices. An inductive coding approach was employed, in which narrative responses were first reviewed to identify recurring concepts, terminology, and decision rationales expressed by participants \cite{Braun_Clarke_ThematicAnalysis_2006}. These initial codes were then grouped into higher-level themes aligned with the study's research questions, such as reliance on assessor judgment, risk-based sampling, control criticality, environmental complexity, and perceived gaps in guidance. To enhance analytical rigor, coding decisions were applied consistently across responses and iteratively refined to ensure that themes accurately represented the underlying data while preserving contextual meaning.

\subsection{Integration of Findings}
Findings across all survey sections were integrated to develop a coherent and holistic understanding of evidence sampling practices within the CMMC assessment ecosystem. Results from participant background data, structured survey items, scenario-based questions, and open-ended responses were examined collectively to identify patterns of convergence and divergence in assessor behavior and perceptions. Quantitative distributions were used to contextualize and corroborate qualitative themes, while narrative explanations provided depth and explanatory context for observed variability. This integrative approach enabled triangulation across data types, strengthened the validity of conclusions, and ensured that findings reflected systemic trends rather than isolated responses.

\subsection{Interpretation of Findings}
The final step of the research focused on interpreting the integrated findings in relation to established audit theory and the governance structure of the CMMC program. Identified themes and patterns were examined against core audit principles such as representativeness, consistency, traceability, and professional judgment to assess how current evidence sampling practices align, or fail to align, with recognized assessment standards. These findings were further contextualized within the CMMC governance model, including assessor certification requirements, C3PAO oversight, and the objectives of uniform certification outcomes. 

\section{Results \& Discussion}

\subsection{Participant Background}

Analysis of participant background data indicates that respondents represent an experienced subset of the CMMC assessor population. All respondents identified as either CCAs or LCCAs. Most participants reported multiple years of cybersecurity and assessment experience (Table \ref{tab:YearsExperience}) with most (94\%) indicating prior involvement in more than two CMMC assessment engagement (Table \ref{tab:AssessmentsAmount}). This experience profile establishes that responses reflect informed professional judgment rather than novice interpretation.

Despite variation in years of experience and the number of assessments performed, no clear clustering was observed that would suggest a dominant assessor profile. Instead, the participant pool reflects diversity in assessment exposure, organizational context, and assessor roles. This diversity provides an appropriate foundation for examining variability in evidence sampling practices across practitioners.

\begin{table}[ht]
    \centering
\caption{Respondents' Cybersecurity/Audit Experience}
\label{tab:YearsExperience}
    \begin{tabular}{lcc}
        \toprule
        Years of Experience& Count(n) & Percent(\%)\\ 
        \midrule
        0--2 years       & 0 & 0\\
        3--5 years       & 5 & 31\\
        6--10 years      & 2 & 13 \\
        More than 10 years & 9 & 56 \\
        \bottomrule
    \end{tabular}
\end{table}

\begin{table}[ht]
    \centering
\caption{Number of CMMC Assessments Previously Performed by Respondents}
\label{tab:AssessmentsAmount}
    \begin{tabular}{lcc}
        \toprule
        Assessments Performed & Count(n) & Percent(\%)\\
        \midrule
        1--2        & 1 & 6\\
        3--5        & 5  & 29\\
        6--10       & 1  & 6\\
        More than 10 & 10 & 59\\
        \bottomrule
    \end{tabular}
\end{table}

\subsection{Evidence Sampling Practices}

Most respondents (71\%) reported that their C3PAO does not provide an evidence sampling methodology (Table \ref{tab:SamplingProvideByC3PAO}), suggesting sampling approaches are often developed at the assessor or assessment-team level rather than driven by a uniform internal standard. In addition, respondents perceived limited evidence-sampling guidance in existing standards: 44\% indicated no formal guidance and 31\% reported only general guidelines/principles, while 13\% characterized the guidance as highly detailed/prescriptive (Table \ref{tab:C3PAOGuidanceLevel}). When no C3PAO sampling standard exists, respondents reported using a range of sampling approaches (see Table \ref{tab:EvidenceApproachNotC3PAOProvided}).

\begin{table}[ht]
    \centering
\caption{Whether Respondents' C3PAOs Provide an Evidence Sampling Methodology}
\label{tab:SamplingProvideByC3PAO}
    \begin{tabular}{lcc}
        \toprule
        Response & Count(n) & Percent(\%)\\
        \midrule
        Yes & 5 & 29\\
        No  & 12 & 71\\
        \bottomrule
    \end{tabular}
\end{table}

\begin{table}[ht]
    \centering
\caption{Perceived Level of Evidence-Sampling Guidance in Existing Standards}
\label{tab:C3PAOGuidanceLevel}
    \begin{tabular}{l cc}
        \toprule
        Level of Guidance& Count(n) & Percent(\%)\\ 
        \midrule
        No formal guidance& 7 & 44\\
        Minimal or unclear guidance& 2 & 13\\
        General guideline or principles only& 5 & 31\\
        Highly detailed or prescriptive & 2 & 13\\
        \bottomrule
    \end{tabular}
\end{table}

Table \ref{tab:DetermineEvidenceSampled} summarizes the decision factors respondents cited when determining evidence sampling scope. The most frequently cited factors were environment size/complexity (82\%), assessor experience/judgment (76\%), and risk or control criticality (71\%). Only a small minority reported following a quantitative or statistical model (12\%), or other factors (12\%). 

\begin{table}[ht]
    \centering
\caption{Sampling Approaches Used When No C3PAO Sampling Standard Exists (Respondents Reporting ``No'' in Table \ref{tab:SamplingProvideByC3PAO})}
\label{tab:EvidenceApproachNotC3PAOProvided}
    \begin{tabular}{lcc}
        \toprule
        Sampling Approach Used& Count(n) & Percent(\%)\\
        \midrule
        Created by individual assessors 
        & 6 &50\\
        Developed collaboratively by assessment team& 4 &33\\
        Adapted from external frameworks& 1 &8\\
        Other& 4 & 33\\
        \bottomrule
    \end{tabular}
\end{table}

\begin{table}[ht]
    \centering
\caption{Factors Cited for Determining Evidence Sampling Scope. Multiple selections/mentions allowed.}
\label{tab:DetermineEvidenceSampled}
    \begin{tabular}{lcc}
        \toprule
        Decision Factor& Count(n) & Percent(\%)\\ 
        \midrule
        Based on risk or control criticality& 12&71\\
        Based on size or complexity of the environment& 14&82\\
        Based on experience/judgement& 13&76\\
        Following a quantitative or statistical model& 2&12\\
        Other& 2&12\\
        \bottomrule
    \end{tabular}
\end{table}

\subsection{Sampling Scenarios}

Three scenario-based questions were used to assess how assessors apply evidence sampling principles under controlled conditions. Summaries of the open-ended responses are provided. 

\subsubsection{Scenario 1}
Participants were asked, \textit{``A medium-sized organization (200 users, 4 locations) claims compliance for all access control requirements. How much evidence (percentage of total user accounts or systems) would you typically sample to validate compliance?''} Assessor-reported sampling varied dramatically from minimal random checks (e.g., 1-3\%) to near-census review (95-100\%).  Several respondents suggested ~10-20\% as a baseline (sometimes capped at 25\%). Several respondents avoided a flat percentage entirely, describing coverage-based checks instead. Many also conditioned sampling on evidence quality and emphasized stratified sampling across privilege levels, authentication methods, systems, and locations rather than a single flat percentage.

\subsubsection{Scenario 2}
Participants were asked, \textit{``For a high-risk control area (e.g., incident response or audit logging), what level of sampling would you consider sufficient to gain assurance?''} (Table \ref{tab:Scenario2Data}) and then asked to further clarify any `Depends' response. Nearly half of respondents (47\%) selected `Depends,' and their explanations consistently reframed sufficiency as contextual rather than percentage-based, often prioritizing representative exemplars, coverage across log sources/system types, and evidence of sustained operation (e.g., year-long records), while noting that `100\%' can be realistic when the underlying population is small (e.g., annual tabletop tests).

\begin{table}[ht]
    \centering
\caption{Scenario 2: Reported Sampling Levels Considered Sufficient for High-Risk Controls}
\label{tab:Scenario2Data}
    \begin{tabular}{lcc}
        \toprule
        Reported Sampling Level & Count(n) & Percent(\%) \\
        \midrule
        $<$ 10\% & 2 &12\\
        10 - 25\%             & 5  &29\\
        26 - 50\%            & 2 & 12\\
        $>$ 50\%               & 0  &0\\
       Depends (context specific)             & 8  &47\\
        \bottomrule
    \end{tabular}
\end{table}

\subsubsection{Scenario 3}
Respondents were asked, \textit{``When assessing documentation and records (e.g., training records, vulnerability scans), how do you determine sample size adequacy?''} Assessors described `adequate' record sampling using markedly different heuristics, from minimal `one example' standards to 100\% checks for certain populations (notably training) and expectations of exhaustive documentation. Common decision rules included scaling to population size, combining random and mixed selection, ensuring time-period coverage (e.g., one year/cycle), and evaluating representativeness (role/system coverage) rather than raw volume.

\subsection{Experience and Consistency}

Reported frequencies of sampling inconsistencies observed between assessors during assessments are shown in Table \ref{tab:SamplingInconsistancies}.  Whether respondents attribute assessment issues to unclear or inconsistent evidence sampling practices is summarized in Table \ref{tab:InconsistantSamplingIssues}. Importantly, inconsistency was reported by both less experienced and more experienced assessors, indicating that variability is not confined to early-career practitioners. Narrative responses suggest that inconsistency is a recurring operational issue rather than an isolated occurrence.

\begin{table}[ht]
    \centering
\caption{Frequency of Sampling Inconsistencies Observed Between Assessors During Assessments}
\label{tab:SamplingInconsistancies}
    \begin{tabular}{l cc}
        \toprule
        Reported Frequency & Count(n) & Percent(\%)\\ 
        \midrule
        Very frequently & 3 &18\\
        Occasionally    & 9 &53\\
        Rarely          & 2 &12\\
        Never           & 3 &18\\
        \bottomrule
    \end{tabular}
\end{table}

\begin{table}[ht]
    \centering
\caption{Whether Respondents Attribute Assessment Issues to Unclear or Inconsistent Evidence Sampling Practices}
\label{tab:InconsistantSamplingIssues}
    \begin{tabular}{lcc}
        \toprule
        Response & Count(n) & Percent(\%) \\ 
        \midrule
        Yes& 5 &29\\
        No& 11 &65\\
        Unsure& 1 &6\\
        \bottomrule
    \end{tabular}
\end{table}
 
\subsection{Recommendations and Perspectives}

When asked whether standardized guidance for evidence sampling would improve CMMC assessments, most respondents expressed support for some form of standardization. However, respondents generally opposed rigid, percentage-based requirements. Instead, they advocated for structured, risk based guidance that establishes minimum expectations while preserving assessor discretion.

Common recommendations included the development of baseline sampling criteria, clearer definitions of evidentiary sufficiency, improved assessor training and calibration, and alignment with established audit and assessment standards. These perspectives indicate broad practitioner recognition of the limitations of the current approach and openness to methodological enhancement.

 \section{Summary of Results}

\subsection{RQ1 (Decision drivers)}
Open-ended responses revealed that sampling decisions are most frequently influenced by perceived risk, control criticality, and environmental complexity. Respondents often described adjusting sample size based on system scope, user population, or perceived impact of control failure. Formal statistical or percentage-based sampling models were rarely referenced, suggesting that evidence sampling is primarily heuristic rather than formulaic.

\subsection{RQ2 (Dispersion under identical scenarios)}
Results indicate substantial variability in assessor responses across all scenarios presented. In scenarios involving moderately sized environments, proposed sampling levels ranged widely, with no clear consensus on minimum or sufficient thresholds. For scenarios involving higher-risk controls, respondents generally indicated a need for increased sampling; however, the magnitude of that increase varied significantly. In documentation and record based scenarios, respondents differed in their interpretation of what constituted adequate evidence, with some favoring representative sampling and others indicating a preference for exhaustive review. The dispersion of responses across identical scenarios demonstrates that assessors apply materially different sampling approaches, even when contextual variables are held constant. This variability provides empirical evidence of inconsistent application of sampling logic within the current assessment ecosystem. 

\subsection{RQ3 (Experienced inconsistency across assessments)}
A substantial proportion of respondents reported having encountered inconsistencies in evidence sampling expectations across assessments. These inconsistencies were reported both between different assessors and across different C3PAOs.
Even so, respondents broadly support the development of standardized, risk-informed guidance to improve consistency and reliability, while opposing rigid, percentage-based requirements

These findings establish a clear empirical foundation for examining evidence sampling as a methodological gap within the CMMC assessment framework and directly inform the development of a proposed evidence sampling framework and subsequent quantitative validation efforts.

 \section{Limitations}

Several considerations and limitations must be acknowledged when interpreting the findings of this study. Data collection relied on a voluntary, web-based survey distributed to a targeted professional population, which introduces the potential for self-selection bias. Potential participants were identified using publicly available contact information provided by CCAs and (LCCAs within the official CyberAB marketplace. To ensure that survey invitations were delivered only to the intended assessor population, the email list was deliberately refined prior to distribution. Email addresses with generic or organizational suffixes (e.g., info@, cmmc@) were removed to reduce the likelihood of delivery to non-assessor recipients or shared mailboxes. Following this refinement process, 556 survey candidates were identified and invited to participate. Of these, 17 participants submitted usable responses, resulting in a low response rate that must be considered when interpreting the results.

Given the small number of respondents, the qualitative portion of the study may not have achieved thematic saturation across all topics \cite{Hennink_Kaiser_Saturation_2022_b7}. The limited number of responses may constrain the generalizability of the findings and preclude the application of advanced inferential statistical techniques. However, this limitation should be viewed in the context of the highly specialized nature of the target population and the professional sensitivity associated with CMMC assessments. The respondents represent experienced practitioners, and the convergence of themes across multiple survey sections and scenario-based questions suggests that the findings reflect systemic patterns rather than isolated perspectives. Accordingly, the study is positioned as an exploratory and diagnostic investigation intended to identify methodological gaps within CMMC evidence sampling practices, rather than to produce statistically representative conclusions.

Additional limitations arise from the reliance on self-reported data rather than direct observation of assessment activities. While anonymity was intentionally implemented to promote candid disclosure, without the collection of personally identifiable information or IP address data, responses reflect participants' reported practices and perceptions. To mitigate this limitation, the survey emphasized operationally grounded questions and structured sampling scenarios that required applied decision-making. Finally, thematic coding of open-ended responses was conducted systematically and iteratively by a single researcher and did not include formal inter-rater reliability testing. Future research should incorporate independent coders and quantitative reliability measures to further strengthen methodological rigor and validate the findings presented in this study.

\section{Conclusion and Future Work}

The findings of this study identify evidence sampling as a critical methodological gap within the current CMMC assessment framework; however, they also establish a clear foundation for continued research and methodological development. Future work should focus on expanding empirical validation, refining assessment tools, and translating research findings into actionable governance and practice improvements within the CMMC ecosystem.

A large-scale quantitative validation study is recommended to assess the generalizability of the findings identified in this exploratory research. Building on the utilized survey instrument, future research should deploy a validated Likert-scale questionnaire to a broader population of CCAs and LCCAs. Such a study would enable statistical testing of inter-assessor variability, perceived guidance sufficiency, and support for standardized sampling frameworks. Reliability testing should be employed to confirm construct validity and measurement consistency, while regression modeling may be used to examine relationships between assessor experience, environmental complexity, and sampling decisions.

A proposed CMMC Evidence Sampling Framework (CESF) should be refined and empirically evaluated through pilot implementation. Future work could involve applying the framework in simulated or live assessment environments to compare sampling consistency, assessment efficiency, and outcome alignment against current practice. Comparative analysis between framework-guided and unguided assessments would provide evidence of CESF's effectiveness in improving repeatability and transparency. Results from such pilots could inform iterative refinement of the framework and support its adoption within assessor training and certification programs.

Finally, future research should examine the governance and policy implications of evidence sampling standardization within CMMC. This includes evaluating how standardized sampling guidance could be integrated into CAICO approved assessor training curricula, C3PAO quality management systems, and assessment process documentation. Longitudinal studies assessing the impact of standardized sampling on assessment outcomes, appeals, and stakeholder confidence would contribute to understanding the role of evidence sampling in maintaining the integrity and credibility of the CMMC certification program. Together, these future research directions provide a structured pathway for advancing both 
scholarly understanding and practical implementation of evidence sampling methodologies in cybersecurity certification assessments.

\section*{Acknowledgment}

The authors thank the CCAs and LCCAs who voluntarily participated in this study and shared their professional experiences. Their insights were essential to examining evidence sampling practices within the CMMC assessment ecosystem. The authors also acknowledge informal academic and professional discussions that contributed to the development of the research design and analysis. No external funding was received for this research. 

\balance
\printbibliography

\end{document}